# Ultraviolet Raman Microscopy of Single and Multi-layer Graphene

Irene Calizo, Igor Bejenari[+], Muhammad Rahman, Guanxiong Liu and Alexander A. Balandin[*]

Nano-Device Laboratory, Department of Electrical Engineering and Materials Science and Engineering Program, Bourns College of Engineering, University of California – Riverside, Riverside, California 92521 U.S.A.

*(Complete version of the paper was submitted for journal publication on February 20, 2009)*

## Abstract

We investigated Raman spectra of single-layer and multi-layer graphene under ultraviolet laser excitation at the wavelength $\lambda$=325 nm. It was found that while graphene's *G* peak remains pronounced in UV Raman spectra, the *2D*-band intensity undergoes severe *quenching*. The evolution of the ratio of the intensities of the *G* and *2D* peaks, *I(G)/I(2D)*, as the number of graphene layers *n* changes from *n*=1 to *n*=5, is different in UV Raman spectra from that in conventional visible Raman spectra excited at the 488-nm and 633-nm wavelengths. The *2D* band under UV excitation shifts to larger wave numbers and is found near 2825 cm$^{-1}$. The observed UV Raman features of graphene were explained by invoking the resonant scattering model. The obtained results contribute to the Raman nanometrology of graphene by providing an additional metric for determining the number of graphene layers and assessing its quality.

---

+On leave from the Institute of Electronic Engineering and Industrial Technologies of the Academy of Sciences of Moldova, Chisinau, Republic of Moldova

[*] Corresponding author; electronic address (A.A. Balandin): balandin@ee.ucr.edu

Irene Calizo, Igor Bejenari, Muhammad Rahman, Guanxiong Liu and Alexander A. Balandin, UCR, 2009

Since its recent mechanical exfoliation, graphene attracted tremendous attention of the scientific and engineering communities [1-2]. Graphene, which consists of a single atomic layer of hexagonally arranged carbon atoms, revealed a number of unique properties including extremely high electron mobility reaching up to ~200000 cm$^2$/Vs [3-4]. Another useful property of graphene is its extremely high thermal conductivity exceeding ~3000 W/mK near room temperature (RT) [5-6]. The latter is beneficial for proposed graphene devices and suggests a possibility of thermal management applications in electronics and optoelectronics [6]. Owing to its properties, graphene emerged as a promising material for electronic [1-4, 7] and spintronic devices [8-9], low-noise transistors [10-11], interconnects [12], extremely sensitive detectors [13], and heat removal [6].

Graphene is produced by variety of techniques including mechanical exfoliation [1-6], growth on the basis of SiC substrates [14], high-pressure high-temperature (HPHT) growth [15], chemical exfoliation [16] and electrostatic deposition [17]. In all of the graphene exfoliation or synthesis techniques one needs to accurately determine the number of atomic planes in the obtained samples (here and below the term "atomic plane" is used interchangeably with the "atomic layer"). Indeed, graphitic samples with the number of atomic planes above approximately 7-10 can hardly be considered a two-dimensional (2-D) material system and referred to as "graphene". Ensuring the quality of the resulting material, e.g. low defect density and crystallinity, is also important for any study of graphene properties and graphene device applications. Micro-Raman spectroscopy has emerged as a reliable, non-destructive and high-throughput technique for determining the number of atomic layers and for graphene quality control [18-22]. There are very few alternatives to this technique, e.g. low-temperature transport measurements or cross-sectional transmission electron microscopy (TEM), but all of them are much harder to perform and most of them are destructive.

In addition to being a nanometrology tool for graphene, Raman spectroscopy provides a wealth of information about the physical and chemical properties of the material. It has been commonly used for the characterization of carbon materials [23-24] and played a key role in the study of graphene's properties such as electron – phonon interactions [25], lattice inharmonicity [21], defects and phase transitions [26]. To date, the Raman spectroscopy of graphene has been limited to the visible (VIS) optical frequency range. Conventionally, the spectra are excited by VIS lasers, commonly, with the wavelength 488 nm, 514 nm and 633nm, and detected in the backscattering configuration. In this letter we report the first study of Raman spectrum of graphene excited by the ultraviolet (UV) laser with the wavelength $\lambda$=325 nm. It is found that the UV excited Raman bands of the single-layer graphene (SLG), bilayer graphene (BLG) and few-layer graphene (FLG) are substantially different from those obtained under VIS excitation. The obtained UV Raman information can be used as an additional metric in Raman nanometrology of graphene. It also sheds new light on the resonant electron – phonon interaction in graphene.

The informative Raman bands of graphene, excited by VIS lasers, are in the range from ~1000 $cm^{-1}$ to 3500 $cm^{-1}$. The most pronounced features are a distinctive peak in the vicinity of 1580 $cm^{-1}$ and a band around 2700 $cm^{-1}$. The first-order $G$ peak at 1580 $cm^{-1}$ corresponds to $E_{2g}$-symmetry phonons at the BZ center. The second-order band near 2700 $cm^{-1}$ has been referred to as *2D* band in the context of graphene research (in the former carbon materials nomenclature it was called *G'* band). The presence of the *2D* band in the Raman spectra of graphene and its sensitivity to the number of graphene atomic layers has been explained on the basis of the double resonance model [18]. The disorder-induced $D$ peak appears in graphene spectrum at ~1360 $cm^{-1}$ when the sample has a large concentration of defects.

The shape of the *2D* band and ratio of the intensity of the $G$ peak and *2D* band, denoted as *I(G)/I(2D)*, has been used for identification of the number of atomic graphene layers $n$ from $n$=1 to approximately $n$=7. The further increase in the number of layers leads to a Raman spectrum identical to that of bulk graphite. The double-resonance model, which predicts the number and position of the elemental peaks within the *2D* band, helps in determining the number of atomic planes from the experimentally measured *2D* band, and explains the band's shape sensitivity to the number of graphene atomic layers. The deconvolution of the 2D into elemental peaks has made it possible to determine the number of atomic planes even when graphene is placed on non-standard (other than Si/$SiO_2$) substrates with many defects [27-28]. The change in the ratio *I(G)/I(2D)* with the increasing number of layers $n$ was determined empirically, and used as an independent metric for the verification of the number of layers.

It is rather common that due to resonance enhancement and other effects, the UV-excited Raman spectrum is substantially different from those excited by VIS lasers. The differences are pronounced for diamond and have also been observed in single-wall carbon nanotubes (CNTs) [29]. It was established that VIS Raman spectroscopy is more sensitive to the $sp^2$ carbon sites [30-31] while UV Raman spectroscopy is more sensitive to the $sp^3$-bonded sites. The former was attributed to the fact that VIS excitation resonates with the $\pi$ states of the $sp^2$ sites. For some carbon – based materials, UV Raman is the only method of obtaining informative bands because VIS Raman spectra of such materials are obscured by interference from fluorescence or scattering from $sp^2$-bonded carbon. The T peak at ~1060 $cm^{-1}$ observed in some carbon materials only under UV excitation is due to the C-C bond $sp^3$ vibrations and as such is not expected in UV spectra of $sp^2$-bonded graphene. The expected differences in graphene spectrum under UV and VIS excitation and mechanisms behind them made this UV Raman study of SLG and FLG particularly intriguing.

Graphene samples were prepared by micromechanical cleavage of highly oriented pyrolytic graphite (HOPG) and transferred onto standard oxidized silicon substrates, i.e. $Si/SiO_2$. The thickness of the oxide layer on these substrates has been verified with the spectroscopic ellipsometry and determined to be 308 nm. The Raman spectra were measured at RT in the back scattering configuration with the help of Renishaw Invia micro-Raman spectrometer fitted with VIS (488 nm and 633 nm) and UV (325 nm) lasers. The 1800 lines/mm (VIS) and 2400 lines/mm (UV) gratings were used for recording the spectra. The graphene layers were initially identified by color under an optical microscope fitted with a 100X objective (NA=0.85) and then accurately verified with VIS Raman spectroscopy at $\lambda$=488 nm through the process outline previously [21-22, 27-28]. An optical microscopy image of a typical mechanically exfoliated graphene flake is shown in Figure 1. The regions with different number of atomic layers, e.g. BLG or three-layer graphene (TLG), are indicated by arrows. Our samples were subjected to additional inspection though scanning electron microscopy (SEM) and current – voltage measurements. The samples were kept in a vacuum box between measurements to minimize surface contamination. The excitation power was kept at minimum to avoid local heating and damage to graphene.

The VIS Raman spectra of SLG and FLG are presented in Figure 2. It shows the evolution of the Raman spectrum of graphene excited by the laser light with the wavelength $\lambda$=488 nm as the number of graphene atomic layers changes from $n$=1 to $n$=5. As one can see, graphene has a sharp single *2D* peak, which transforms to broader bands consisting of several elemental peaks as the number of atomic layers $n$ increases. The spectral position of the peaks, their evolution with the increasing number of graphene layers and changes in *I(G)/I(2D)* ratio are consistent with previous VIS Raman studies of graphene [18-22]. This confirms that the number of atomic layers was determined correctly and indicates that our results are easily reproducible. The disorder-induced *D* peak is completely absent in all spectra excited at $\lambda$=488 nm. The VIS Raman spectra presented in Figure 2 and the absence of *D* peak attest to the high quality of the examined SLG and FLG.

The UV Raman spectra were excited at $\lambda$=325 nm using low power density on the surface to avoid damage to the graphene lattice. In Figure 3 we present UV Raman spectrum of graphene as the number of atomic layers $n$ changes from $n$=1 to $n$=5. All spectra have excellent signal-to-noise ratio. Only the spectrum of SLG reveals a weak disorder-related *D* peak. The latter is likely related to the UV-laser damage during the measurement and suggests that the damage threshold for SLG is lower than that for FLG. The *G* peak on all spectra appears at the same position at 1580 cm$^{-1}$ and its intensity growth in a similar way to that in VIS spectrum as $n$ increases. This trend is related to the growing interaction volume for Raman scattering as the number of layers increases. The *G* peak is associated with BZ center phonons, which explains why its position does not change with the excitation wavelength. The striking difference between UV Raman spectra of graphene and its VIS Raman spectra is severe *quenching* of the 2D band intensity in UV Raman spectra for all number of graphene layers $n$.

The intensity ratio *I(G)/I(2D)* for SLG in VIS Raman spectrum is about 0.24 under the 488-nm excitation. This ratio becomes *I(G)/I(2D)* ~ 9 for UV Raman spectrum of SLG and grows even farther to *I(G)/I(2D)* ~ 23 for bilayer graphene. Table I summarizes the *I(G)/I(2D)* ratios for VIS ($\lambda$=488 nm) and UV Raman spectra of graphene as the number of atomic layers changes from $n$=1 to $n$=5. The different dependence of *I(G)/I(2D)* ratio on the number of layers in UV Raman can be useful for the spectroscopic Raman nanometrology of graphene by helping to accurately determine $n$. It provides an additional metric for counting the number of atomic planes to the currently used shape of *2D* band and *I(G)/I(2D)* in VIS Raman.

Although the *2D* band in UV Raman is strongly suppressed one can still extract the spectral position of its maximum through Lorentzian fitting. The position of the *2D*-band maximum depends strongly on the excitation wavelength. The *2D* band under UV excitation shifts to the larger wave numbers and is found near 2825 cm$^{-1}$. This is a large shift of about 185 cm$^{-1}$ as compared to the band position at ~2640 cm$^{-1}$ found under 633-nm excitation. The quantitative analysis of the excitation energy dependence is given below in the framework of the resonant Raman scattering model.

We consider the *2D* band in the framework of the double-resonant or fully resonant processes, which take place along the $\Gamma - K - M - K' - \Gamma$ direction in the first BZ [32-36]. The electron – phonon coupling is the strongest for transverse optical (TO) phonons and electrons near the K and K' points [18, 36]. The schematics for possible resonant interactions are presented in Figure 4 (a-b). The appearance of the first-order $G$ peak is explained in the following way. An electron excited by the laser irradiation with an energy $E_{ex}$ makes a transition from the initial state $\boldsymbol{o}$ vertically to the state $\boldsymbol{a}$ in the conduction band. After emitting a phonon with a close-to-zero momentum, the electron transfers to the virtual state $\boldsymbol{i}$. Recombination with a hole in the same point of the BZ results in the emission of a photon with the energy $E_{ex} - \hbar\omega_{ph}(0)$.

For the description of the *2D* peak one can assume the following Stokes scattering process. An incident photon with the energy $E_{ex}$ creates an electron–hole pair with the quasi-momentum $\boldsymbol{k}_0$ within the interval $\Gamma - K$. The electron scatters to the state $\boldsymbol{b}$, characterized by the quasi-momentum $\boldsymbol{k}_b$ through emission of a phonon with the momentum $\boldsymbol{q} = \boldsymbol{k}_b - \boldsymbol{k}_0$ [18, 36]. From this point, according to Basko [35], the scattering proceeds either through the fully resonant channel or double resonant channel. In the former process, the electron and hole can recombine at point $\boldsymbol{b}$ with the emission of a photon (see Figure 4 (a)), which contributes to the two-phonon *2D* Raman peak [35]. In the fully resonant process, the scattering involving a hole is similar but with the momentum of $-\boldsymbol{q}$. In the double resonant process, an electron, which does not recombine with the hole at point $\boldsymbol{b}$, is scattered to the virtual state $\boldsymbol{i}$ (see Figure 4 (b)) through emission of another phonon with the opposite quasi-momentum $-(k_b\text{-}k_0)$ and energy $\hbar\omega_{ph}(k_b - k_0)$ [36]. It is followed by the electron – hole recombination with the emission of a photon, which contributes to the *2D* Raman band.

For both considered channels (Figure 4 (a-b), the energy conservation condition for the first transition can be written as

$$E_{ex}(\boldsymbol{k_o}) = \left[E_c(\boldsymbol{k_b}) - E_v(\boldsymbol{k_o})\right] + \hbar\omega(\boldsymbol{k_b} - \boldsymbol{k_o}). \qquad (1)$$

Here $E_{ex}$ is the laser excitation, $\hbar\omega_{ph}$ is the energy of a phonon emitted during the electron transition from the state $\boldsymbol{k_o}$ to $\boldsymbol{k_b}$. Other notations correspond to those in Figure 4 (a-b). We assume that the electron (hole) energy dispersion $E_v(\boldsymbol{k_o})$ and $E_c(\boldsymbol{k_b})$ is given by the Slatter – Koster expression [33]

$$E(\boldsymbol{k}) = \pm\gamma_0\sqrt{1 + 4\cos^2\left(\frac{\sqrt{3}}{2}a_c k_x\right) + 4\cos\left(\frac{\sqrt{3}}{2}a_c k_x\right)\cos\left(\frac{3}{2}a_c k_y\right)}, \qquad (2)$$

where, $a_c$=1.422 A is the inter-atomic spacing and $\gamma_0$=−2.7 eV is the inter-atomic overlap energy in graphene. We use a conventional linear interpolation for the TO phonon frequency along the K – M direction in the form $\omega_{ph}(q_x) = 630\sqrt{3}a_c q_x/\pi + 850$ cm$^{-1}$. In this interval the wave vector components are related as $q_y = -\sqrt{3}q_x + 4\pi/3a_c$ and K and M coordinates, with respect to the $\Gamma$ point, are K$\left(2\pi/3\sqrt{3}a_c, \quad 2\pi/3a_c\right)$ and M$\left(\pi/\sqrt{3}a_c, \quad \pi/3a_c\right)$.

Solving Eq. (1) and Eq. (2) jointly through iterations, we find numerically the dependence of the *2D*-band position on the excitation energy. The results are the same for both scattering channels. The calculated shift of *2D* band is shown in Figure 5. Our experimental results are also given for comparison. The obtained theoretical dependence for the *2D* band dispersion with excitation energy $E_{ex}$ are rather close to the measurements. While the agreement is excellent in VIS range, the discrepancy between the theory and experiment is larger in UV range. It is interesting to note that the dependence of the *2D*-peak frequency with the excitation energy is superlinear, particularly in UV range. We attributed this feature to contributions from the optical transitions involving electrons with the momentum away from the K-point when the excitation energy is larger than 3 eV. In this region of the BZ the electron dispersion becomes nonlinear.

It is illustrative to compare our results for the dependence of the *2D* band maximum on the excitation energy with available data for graphene and graphite. The slope of our theoretical curve at the higher excitation energy in the UV region is $\delta\omega/\delta E_{exc}$=118 cm$^{-1}$/eV. The experimental shift, which we determined from the three measured points through VIS – UV region, is $\delta\omega/\delta E_{exc}$=92 – 107 cm$^{-1}$/eV. Our experimental data compares well with ~95 – 85 cm$^{-1}$/eV extracted by Casiraghi *et al* [37] for the *2D* band of graphene using two VIS excitation wavelengths $\lambda$=514 nm and $\lambda$=633 nm. The shift for the first-order *D* band calculated by Thomson and Reich [34] for graphite is around 60 cm$^{-1}$/eV while the measured one for graphite is ~ 46 – 51 cm$^{-1}$/eV [38-39]. Taking into account that the dispersion with the excitation energy of the *2D* band has to be approximately double of that for the *D* peak, our results are in line with the previous studies.

In conclusion, we reported the first UV Raman spectra of graphene. It was found that while the *G* peak remains pronounced in UV Raman spectra, the *2D*-band intensity undergoes severe *quenching*. The evolution of the ratio of the intensities of the *G* and *2D* peaks as the number of graphene layers increases is different in UV Raman spectra from that in conventional VIS Raman. The obtained results contribute to the Raman nanometrology of graphene by providing an additional metric for determining the number of graphene layers.

### *Acknowledgements*

This work was supported, in part, by DARPA – SRC through the Focus Center Research Program (FCRP) and its Functional Engineered Nano Architectonics (FENA) Center and Interconnects Focus Center (IFC). A.A.B. acknowledges useful discussions on Raman spectroscopy of graphene with Drs. A.C. Ferrari and D.M. Basko. I.B. acknowledges financial support of the Fulbright Scholar Program.

## FIGURE CAPTIONS

**Figure 1:** Optical microscopy image of typical graphene flakes, with different number of atomic planes, exfoliated mechanically.

**Figure 2:** Evolution of Raman spectrum of graphene under the visible laser excitation at $\lambda$=488 nm as the number of atomic planes increases from $n$=1 to $n$=5. Raman spectra of graphene excited by 488 nm or 514 nm lasers are conventionally used for graphene identification and counting the number of graphene layers.

**Figure 3:** Ultraviolet Raman spectra of graphene shown for single-layer graphene and few-layer graphene as the number of atomic planes increases from $n$=1 to $n$=5. The UV Raman spectra were obtained under 325-nm laser excitation. Note a drastic *quenching* of the intensity of the 2D band.

**Figure 4:** Schematic of the fully resonant (a) and the double resonant (b) scattering mechanisms for the second-order 2D band in the graphene Raman spectrum.

**Figure 5:** Spectral position of the second-order 2D band in Raman spectrum of graphene as a function of the laser excitation energy. The calculated results are shown by the dashed line; the data points indicate the measured values.

Irene Calizo, Igor Bejenari, Muhammad Rahman, Guanxiong Liu and Alexander A. Balandin, UCR, 2009

**Table I:** Measured I(G)/I(2D) Ratio for Single and Few-layer Graphene

| Number of layers, $n$ | 1 | 2 | 3 | 4 | 5 |
|---|---|---|---|---|---|
| VIS ($\lambda$=488 nm) | 0.24 | 0.74 | 1.14 | 1.49 | 2.1 |
| UV ($\lambda$=325 nm) | 8.96 | 22.9 | 24.9 | 22.9 | 35.5 |

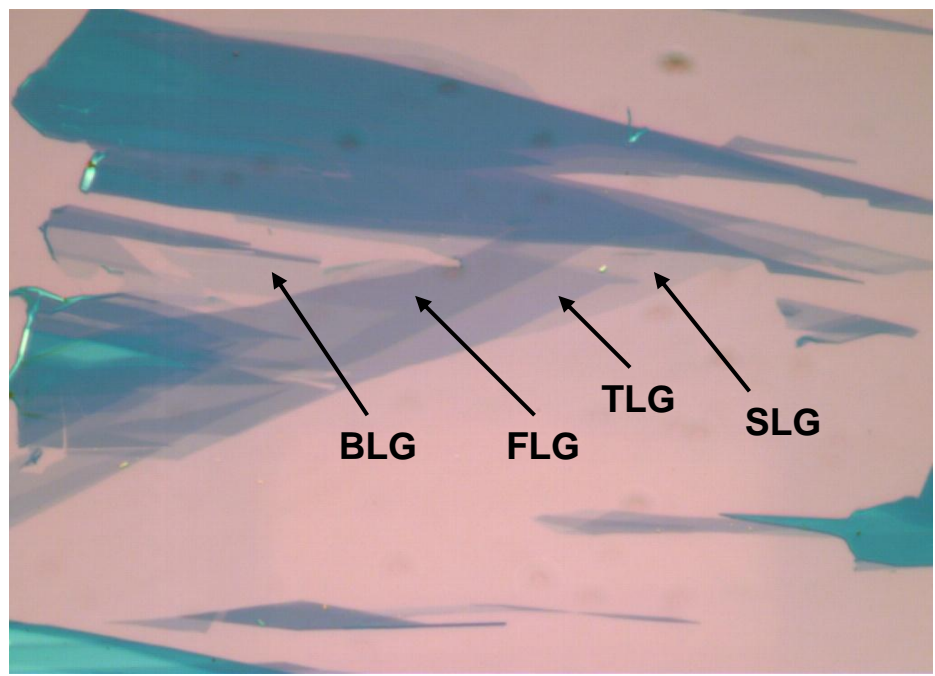

**Figure 1**

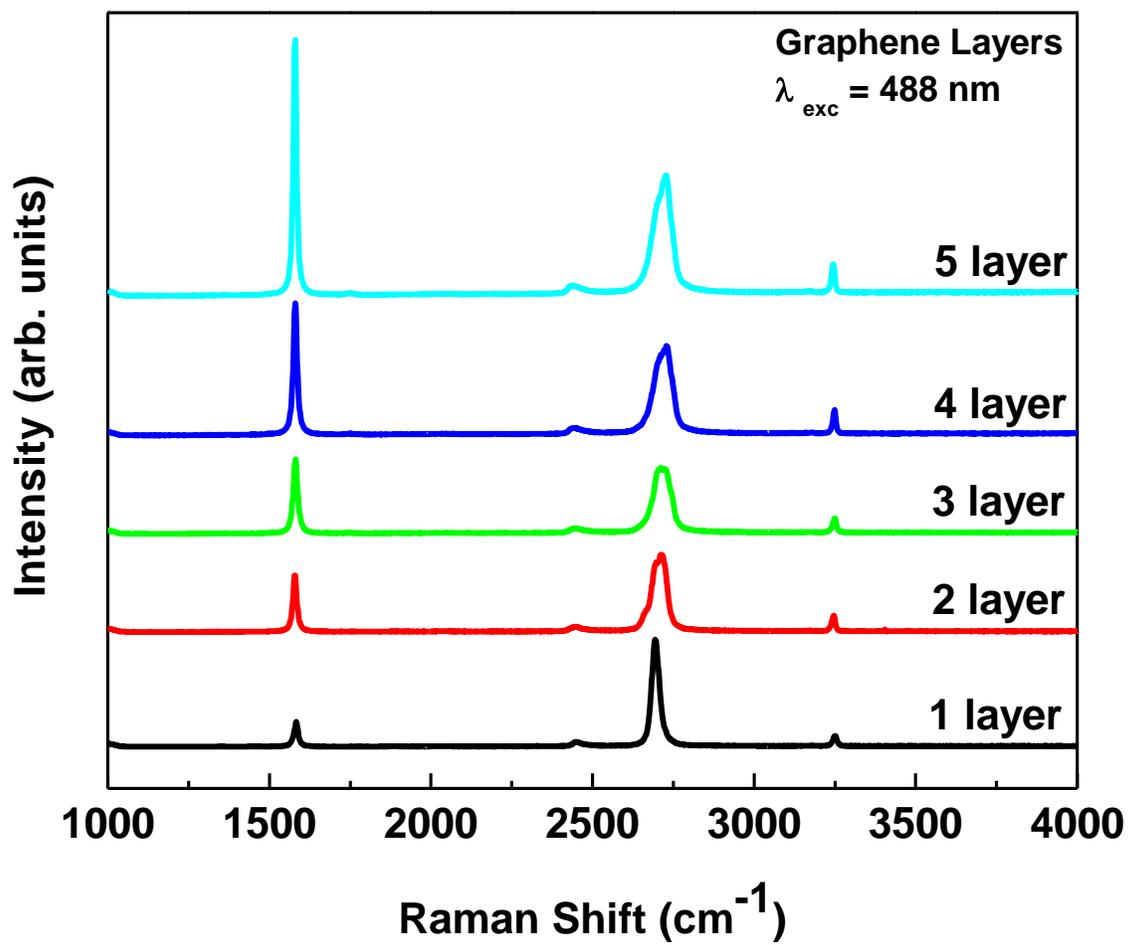

**Figure 2**

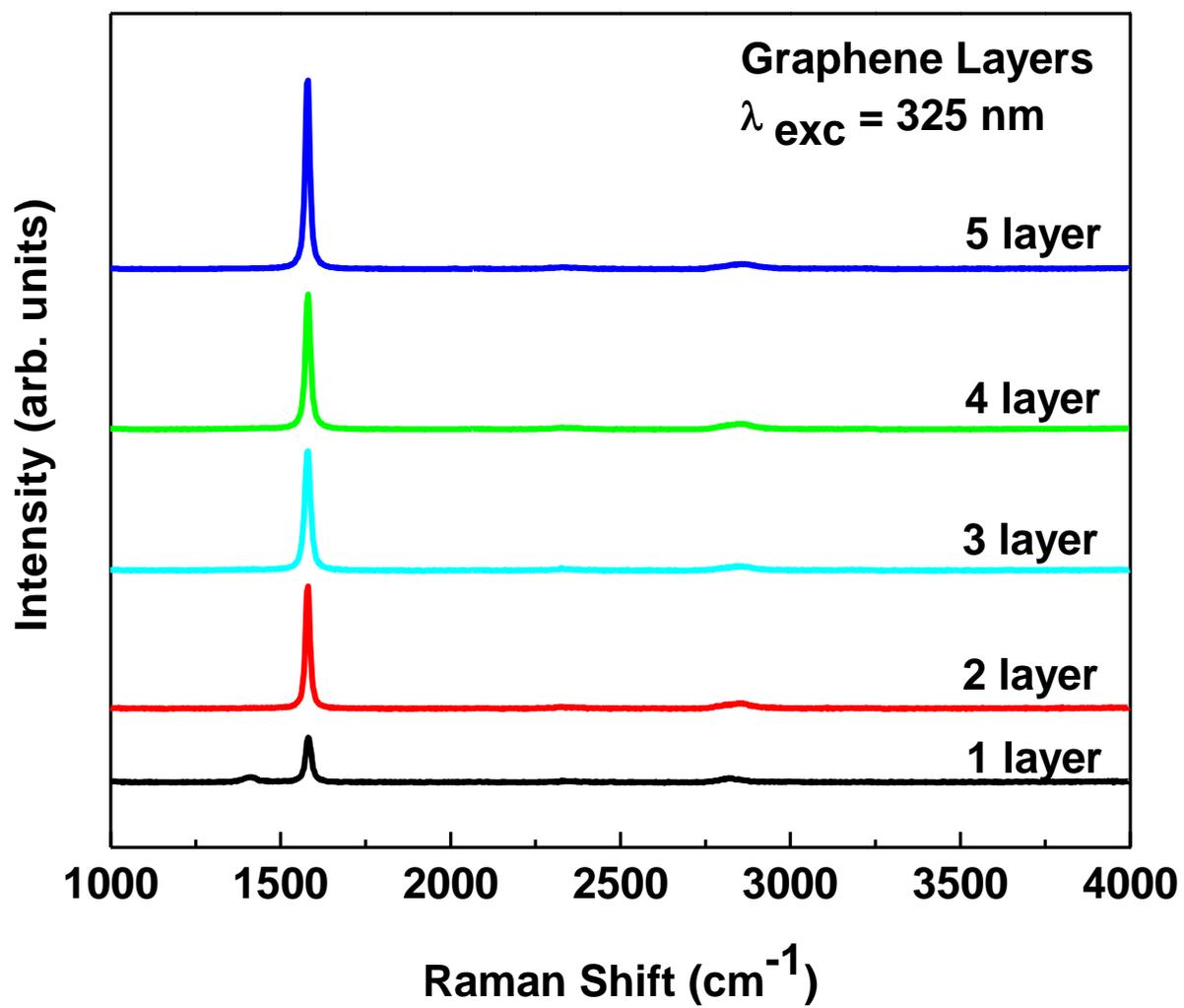

**Figure 3**

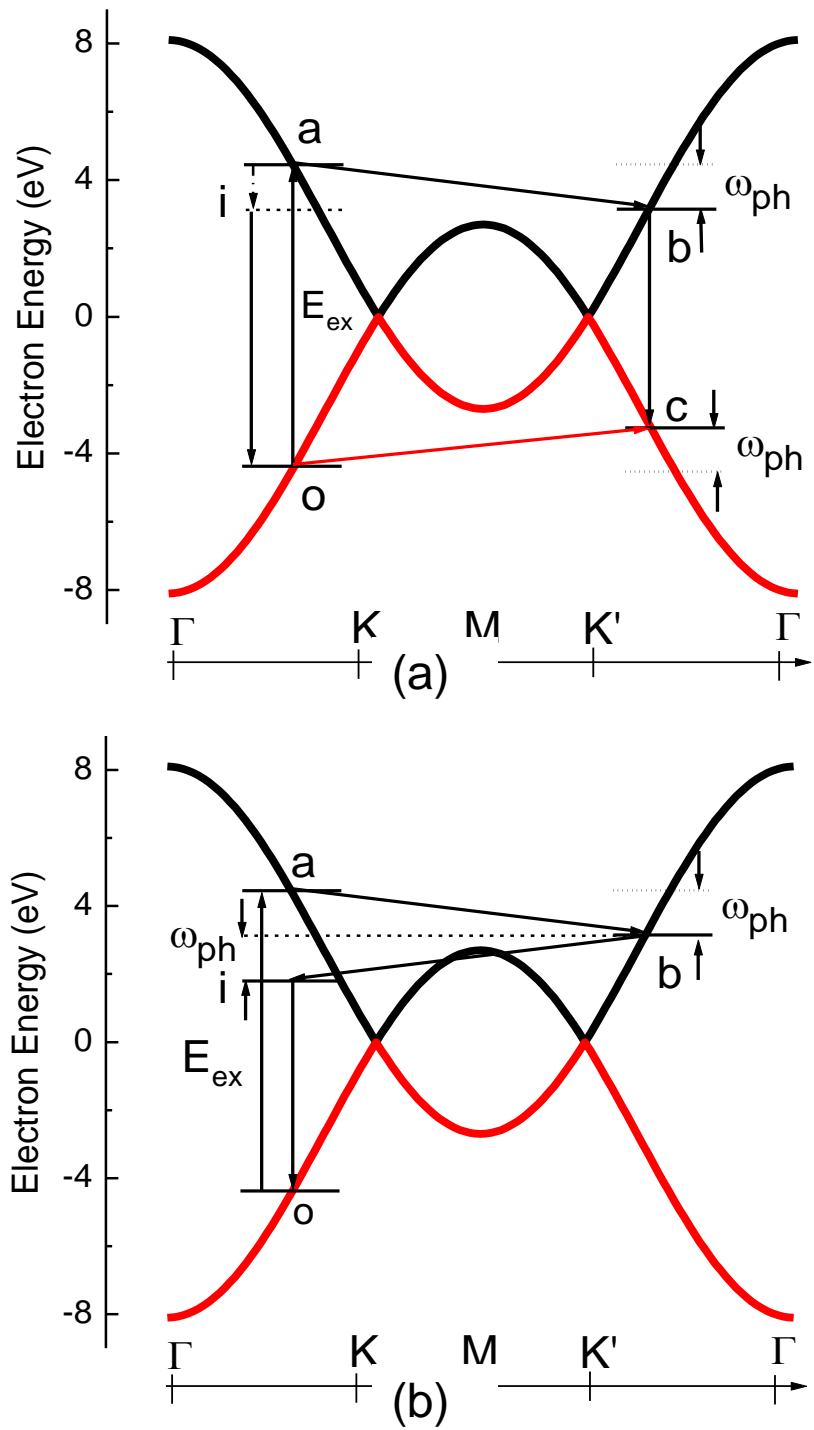

**Figure 4**

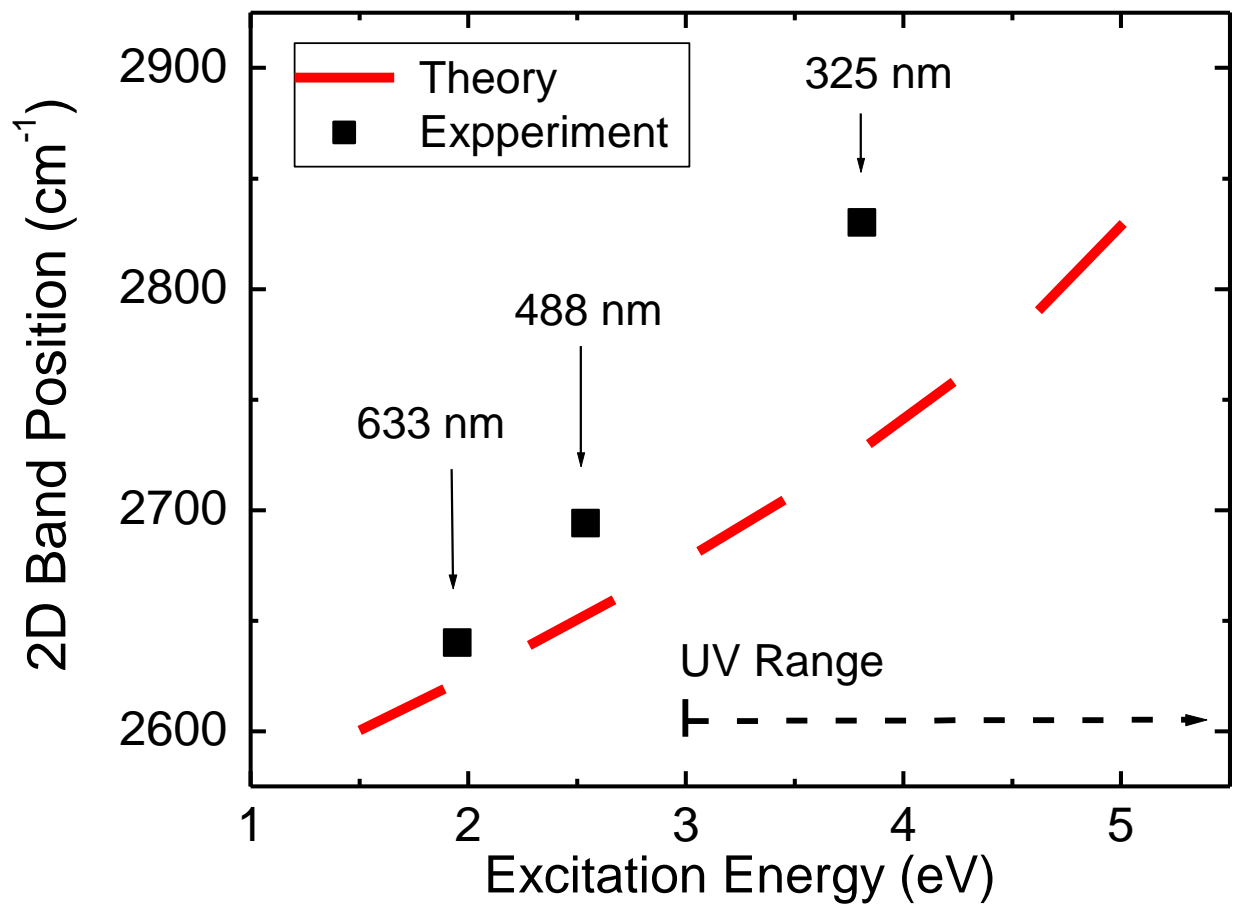

**Figure 5**